\documentclass[aps,prd,twocolumn,nofootinbib,showpacs,superscriptaddress]{revtex4-1}

\usepackage{amssymb}
\usepackage{graphicx}
\usepackage{amsmath, amsthm}
\usepackage{epstopdf}
\usepackage{hyperref}

\newcommand{\be}{\begin{equation}}
\newcommand{\ee}{\end{equation}}

\begin{document}

\title{Do Newtonian large-scale structure simulations fail 
to include relativistic effects?} ñ


\author{Valerio Faraoni}
\email{vfaraoni@ubishops.ca}
\affiliation{Physics Department and STAR Research Cluster, 
Bishop's University,\\
Sherbrooke, Qu\'ebec, Canada J1M~1Z7
}
\author{Marianne Lapierre-L\'eonard}
\email{mlapierre12@ubishops.ca}
\affiliation{Physics Department, Bishop's University,\\
Sherbrooke, Qu\'ebec, Canada J1M~1Z7}
\author{Angus Prain}
\email{aprain@ubishops.ca}
\affiliation{Physics Department and STAR Research Cluster, Bishop's University,\\
Sherbrooke, Qu\'ebec, Canada J1M~1Z7
}

\begin{abstract} 
The Newtonian simulations describing the formation of large-scale structures do not include relativistic effects. A new approach to this problem is proposed, which consists of splitting the Hawking-Hayward quasi-local energy of a closed spacelike 2-surface into a ``Newtonian'' part due to local perturbations and a ``relativistic'' part due to the cosmology. It is found that the Newtonian part dominates over the relativistic one as time evolves, lending support to the validity of Newtonian simulations.  
\end{abstract}



\maketitle

\section{Introduction}
\label{sec:1}

The study of primordial density fluctuations, of their growth 
history, and of the large scale structures that they seed 
is one of the focuses of modern cosmology. Density fluctuations 
and the temperature fluctuations that they induce in the cosmic 
microwave background are one of the foundational  observational tools of theoretical cosmology. The growth of 
large scale structures from their initial seeds depends on the theory of gravity, 
as well as the cosmological parameters and there 
are proposals to test the theory of gravity using cosmology 
\cite{cosmotestGR} (including satellite missions) in addition to studying dark energy and 
early  physics such as baryon acoustic oscillations. Several redshift 
surveys have mapped large scale structures in the sky in the last 
few decades. The observational results are compared with 
theoretical predictions, but there is a potentially very serious 
problem here, which has received proper attention only recently 
\cite{ChisariZaldarriaga, GreenWald, Adamek}.  Most predictions 
for the generation and dynamical evolution of large scale 
structures, which are based on massive $N$-body simulations, are 
done in the Newtonian limit in the dust-dominated era of our 
universe. While they do, of course, account for the  
expansion of the universe, these simulations are 
essentially Newtonian. {\em A priori}, Newtonian physics is appropriate
on small scales, but not on scales comparable to the Hubble 
radius  $H^{-1}$. The increasing computational power has led to a 
corresponding increase in the number of particles $N$ which now 
reaches $ 10^9 - 10^{10}$ and in the size of the box used in these 
simulations, now spanning  $0.5-3$~Gpc. Since at redshift 
$z\simeq 100$ it is $H^{-1}\sim 1.5$~Gpc, the size 
of the box easily exceeds $H^{-1}$ in 
current simulations \cite{ChisariZaldarriaga}. One must therefore 
worry about including relativistic effects in these simulations. 
These include special-relativistic (velocity) and 
general-relativistic (cosmological) effects. 
Naively, one would like to superpose 
the purely local Newtonian gravitational field to a purely 
cosmological ``background'' field described  by unperturbed 
Friedmann-Lema\^itre-Robertson-Walker (FLRW)  
space. This is best done using linear perturbation 
theory, if one wants to restrict to first order effects in the 
local potentials. Special-relativistic effects are negligible to 
first order because the peculiar velocities of density 
perturbations are small 
\cite{ChisariZaldarriaga, GreenWald, Adamek}. There is a 
large literature on the effects of the cosmological expansion on 
local systems (see Ref.~\cite{CarreraGiuliniRMP} for a recent 
review), while the effects of local systems on 
cosmology through backreaction have also been the subject of recent 
debates \cite{Buchert}. The recent study in 
\cite{ChisariZaldarriaga} assumes that general 
relativity correctly describes gravity, that velocity perturbations
can be neglected, and that first order post-Newtonian potentials 
$\psi_N, \phi_N$ describe the perturbations to FLRW space. We 
follow the same assumptions here.

There are two ways of dealing with the perturbations of FLRW space:
either one fixes a gauge (as done in \cite{ChisariZaldarriaga} for
this specific problem), or one adopts a gauge-invariant 
formalism (as done in \cite{GreenWald}). The two recent 
studies  of \cite{ChisariZaldarriaga} and \cite{GreenWald} using 
these different approaches essentially agree on the results.  The 
Newtonian simulations do consider the wrong equations for the 
local potential, the density perturbations used in 
the simulations should be corrected, and the initial displacement of 
particles in the simulations is incorrect 
\cite{ChisariZaldarriaga}. However, this does not matter much: 
in the gauge-dependent discussion the missing terms in the 
equations cancel out so that the potential $\phi_{sim}$ 
in the simulations is  
computed correctly even at large scales. As a result, the 
$N$ dark matter particles are  displaced  correctly 
\cite{ChisariZaldarriaga}. 
A physical explanation for this occurrence, given in 
\cite{ChisariZaldarriaga}, is that the scale of the sound 
horizon ({\em i.e.}, the length travelled by dark matter particles 
since the big bang due to their peculiar  velocities) is much less 
than the Hubble radius $ H^{-1}$. However, these 
calculations and interpretation are obtained in the Newtonian 
conformal gauge and are gauge-dependent. The gauge-invariant 
treatment of \cite{GreenWald} computes the correct equations to 
second order in the potentials and includes peculiar motions. Using
a scheme developed to analyze the backreaction of small scale 
inhomogeneities on large scale dynamics, these authors decompose 
the metric in scalar, vector, and tensor perturbations and, by 
analyzing order by order the Einstein equations for the 
gauge-invariant 
quantities, build a dictionary between Newtonian and relativistic 
physics. The dictionary is considerably simplified on small scales 
$ l \ll H^{-1}$. The first order results agree with those of  
Ref.~\cite{ChisariZaldarriaga}. As usual, the disadvantage of the 
gauge-invariant treatment is that it is much heavier and physically
less intuitive and transparent than the description obtained by 
fixing the gauge. In this paper we propose an 
approach which is gauge-invariant yet physically intuitive 
and transparent. There are no miracles, though: the physical  
intuition preserving gauge-invariance is obtained at the price of 
simplifying (perhaps over-simplifying) the problem. However we 
believe that the approach is an interesting alternative and that, 
since all our knowledge of large scale structure is based 
on $N$-body simulations, the limits of validity of Newtonian 
equations in these simulations should be scrutinized very closely 
with a variety of approaches.

\section{A toy model}
\label{sec:2}

Preserving both the advantage of gauge-invariance and that of 
physical transparency and intuition, we propose a toy model 
in general relativity. This toy model is clearly 
oversimplified and unrealistic but it serves as a good 
introduction to our approach to the problem and will be 
generalized to realistic situations in the following 
section.  The toy model is based on a single, 
spherically symmetric perturbation.

\subsection{Preliminaries}

We begin from the post-Newtonian, asymptotically flat,  
approximation of the spacetime metric in isotropic coordinates
\be \label{postNewtonianmetric}
ds^2=  -\left( 1+2\psi_N \right) d\eta^2 + 
\left( 1-2\phi_N \right) \left( dr^2 +r^2 d\Omega_{(2)}^2 \right) 
 \,,
\ee
where $d\Omega_{(2)}^2=d\theta^2 +\sin^2 \theta \, 
d\varphi^2$ is the metric on the unit 2-sphere. Assuming 
a single perturbation with spherical symmetry, it is
\be
\psi_N=\psi_N(r) \,, \;\;\;\; \phi_N=\phi_N(r) 
\ee
(we do not specify the form of the potentials $\phi_N$ and $\psi_N$ 
yet). In general relativity based on the Einstein-Hilbert action, the two potentials coincide, $\psi_N=\phi_N$.

The areal radius\footnote{Recall that the areal radius in a spherically symetric spacetime is that function $R$ for which SO(3) symmetric surfaces have area equal to $4\pi R^2$} of this spacetime is given by
\be \label{arealradius}
R(r)=r\sqrt{ 1-2\phi_N(r)}= r\left( 1-\phi_N(r)\right)+\mathcal{O}\left(\phi_N^2\right)
\ee
and its gradient is 
\be
\nabla_{a}R = \delta_{a r}\left(1-\phi_N-r\phi_N'\right)
\ee
to first order in the potential $\phi_N$.

In general relativity and in the presence of spherical symmetry, 
the physical mass-energy contained inside a sphere of symmetry is the Misner-Sharp-Hernandez (MSH) 
mass $M_\text{MSH}$ \cite{MSH}. 
It is defined in terms 
of the areal radius $R$ of the enclosing sphere by \cite{MSH}
\be \label{MSHdefinition} 
1-\frac{2M_\text{MSH}}{R} \equiv \nabla^c R \nabla_c R\, ,
\ee
which gives, for the line element~(\ref{postNewtonianmetric}), 
\begin{align}
1-\frac{2M_\text{MSH}}{R}&=g^{ab}\nabla_a R \nabla_b R \notag \\
&=\left(1-2\phi_N\right)^{-1}\left(1-\phi_N -r\phi_N'\right)^2 \notag\\
&\simeq 1-2r\phi_N '\, .
\end{align}
Using Eq.~\eqref{arealradius} and again discarding higher order terms, this can be re-expressed as the differential equation
\be
R^2\frac{d\phi_N}{dR}=M_\text{MSH} \, .\label{E:ode}
\ee
Assuming, for sufficiently large $R$, that $M_\text{MSH}\simeq$~constant, the solution to \eqref{E:ode} as a power series in $R^{-1}$ begins as
\be
\phi_N=-\frac{M_\text{MSH}}{R}+\mathcal{O}\left(R^{-2} \right) \, .
\ee
Therefore, for sufficiently large $R$ we have
\be
\phi_N= -\frac{M_\text{MSH}}{r\sqrt{1-2\phi_N}}\simeq -\frac{m}{r}
\ee
where we denote the Misner-Sharp-Hernandez mass simply with $m$
(this change of notation is completely unnecessary here but will avoid confusion in the next section).

\subsection{The spherically symmetric toy model}

The perturbed FLRW metric that we want to consider is 
\begin{align}
 \label{CNgauge}
ds^2&= a^2(\eta) \Big[ -\left( 1+2\psi_N \right) d\eta^2  \notag \\
&\hspace{17mm}\left.+ \left( 1-2\phi_N \right) \left( dr^2 +r^2 d\Omega_{(2)}^2 \right) 
\right]
\end{align}
in the conformal Newtonian (or diagonal) gauge, where we assume that there is only one spherically symmetric perturbation: 
\be
\psi_N=\psi_N(r) \,, \;\;\;\;\;\; 
\phi_N=\phi_N(r) \,.
\ee
Here $\eta$ is the conformal time of the unperturbed FLRW space related to the comoving time $t$ by $dt=ad\eta$. 
The metric (\ref{CNgauge}) is obtained from the 
post-Newtonian metric \eqref{postNewtonianmetric} by means of the conformal
transformation $g_{ab} \rightarrow \tilde{g}_{ab}=\Omega^2 g_{ab}$ 
with conformal factor $\Omega=a(\eta )$.

The idea of our toy model is 
to introduce an effective potential $\Phi_N \equiv - 
\tilde{M}_\text{MSH}/\tilde{R}$ and to  
split it into two contributions which can be compared, one due to 
the spherical perturbation and the other describing the 
cosmological ``background'', as in  
\be
\Phi_N \equiv \Phi_N( \mbox{local}) + \Phi_N (\mbox{cosmological}) 
\,.
\ee
{\em A priori} such an effective potential seems meaningless and 
the decomposition seems highly gauge-dependent, but this is not the 
case, as we show below. The decomposition we will demonstrate comes from a 
geometric property of 
the Misner-Sharp-Hernandez mass. 

Due to this decomposition property, and 
because 
the metric 
(\ref{CNgauge}) is conformal to the post-Newtonian metric 
(\ref{postNewtonianmetric}), 
we are interested in the transformation property of 
the Misner-Sharp-Hernandez mass $M_\text{MSH} \rightarrow 
\tilde{M}_\text{MSH}$ under conformal 
transformations, which is easy to derive (see 
Ref.~\cite{FaraoniVitagliano14})
\be
\tilde{M}_\text{MSH}=\Omega M_\text{MSH} -\frac{R^3}{2\Omega} \nabla^c\Omega
\nabla_c\Omega -R^2 \nabla^c\Omega\nabla_c  R \,.
\ee
It follows that the Misner-Sharp-Hernandez mass of the spacetime~(\ref{CNgauge}) is given by
\begin{align}
\tilde{M}_\text{MSH}&= am +\frac{ \left( 1-2\phi_N\right)^{3/2} a 
{\cal H}^2r^3}{ 2\left( 1+2\psi_N\right)} \notag \\
& = am+\frac{ {\cal H}^2 \tilde{R}^3}{ 2\left( 
1+2\psi_N\right)a^2}  \notag \\
&\simeq am +\frac{ {\cal H}^2 \tilde{R}^3}{ 2a^2}\left( 
1-2\psi_N\right)   \,. \label{E:decompsition}
\end{align}
where ${\cal H}\equiv a'/a$, a prime denotes differentiation 
with respect to the conformal time $\eta $ of the unperturbed FLRW space, and $\tilde{R}=a(\eta) \sqrt{ 1-2\phi_N}\, r =a(\eta) R$ is the areal radius of the metric $\tilde{g}_{ab}$.  In terms of the comoving Hubble parameter $H \equiv 
\dot{a}/a=a'/a^2={\cal H}/a $, we have to first order
\be \label{MSHtoy}
\tilde{M}_\text{MSH}=ma+ \frac{ H^2 \tilde{R}^3}{ 2}\left( 
1-2\psi_N\right) \,.
\ee

The result in Eq.~(\ref{MSHtoy}) exhibits a clear decomposition of the Misner-Sharp-Hernandez mass 
into two parts: we denote the first term on the right hand 
side ``local part'' and the second term ``cosmological part''. This
decomposition into cosmological and local parts is gauge-invariant, as is 
the full quantity $M_\text{MSH}$. We will demonstrate this invariance in a later section when we embed this result in the more general case lacking spherical symmetry. 

 We can now define the ``cosmological  
effective potential''\footnote{On small scales $\tilde{R} \ll 
H^{-1}$, $\Phi_N $ reduces to the  usual Newtonian potential 
$-m/R$, 
where $m$ is the Newtonian mass.}
\be
\Phi_C \equiv  \frac{H^2\tilde{R}^2}{2} \left(1-2\psi_N\right)
\ee
whose order of magnitude is the square of the physical size 
$\tilde{R}$ of the region of interest in units of the Hubble radius 
$H^{-1}$, and the ``local effective potential''
\be
\Phi_N \equiv -\frac{ma}{\tilde{R}} \simeq -\frac{m}{r} 
\,.
\ee
such that the full potential decomposes as
\be
\frac{\tilde{M}_\text{MSH}}{\tilde{R}}=\Phi= \Phi_N+\Phi_C\, .
\ee

The relative importance of the cosmological and local parts of the
Misner-Sharp-Hernandez mass (what one could call the ``degree of 
non-Newtonianity'' of the system), is measured by the ratio
\begin{align}
\alpha \equiv  \left| \frac{\Phi_C}{\Phi_N} \right| 
&= \frac{H^2\tilde{R}^2\left( 1-2\psi_N \right)}{2}\frac{\tilde{R}}{ ma} \nonumber\\
&\simeq \frac{H^2\tilde{R}^3}{ 2ma}  \nonumber\\
& \equiv  \frac{H^2\tilde{R}^3}{ R_\text{Schw}}  \,
\end{align}
where $R_\text{Schw}=2ma $ is the proper Schwarzschild radius of the
spherical perturbation. The ratio $\alpha$ measures how good the Newtonian 
approximation is: it is good if $\alpha \ll 1$ and poor otherwise.

Weakly bound objects are essentially comoving with the 
cosmological ``background'', $\tilde{R} \simeq ar$. In the dust-dominated era of the universe, during which the 
formation of structures takes place, the scale factor evolves as
$a(t)=a_0 t^{2/3}$ and we have for weakly bound objects that
\be
H^2R^3 \sim  \frac{4}{9} \, a_0^3
\ee
is constant. During this time, the proper Schwarzschild radius increases with the scale factor like $t^{2/3}$. Hence
\be
\alpha \sim \frac{2}{9} \frac{r^3}{m}\, \frac{1}{t^{2/3}}
\ee
decreases.

The conclusion achieved
with our toy model is that,  even if 
numerical simulations of 
structure formation start on scales at which 
Newtonian physics is not adequate, the situation improves with
time. For weakly bound objects, local physics 
comes to dominate over the relativistic cosmic expansion and this
conclusion is gauge-independent but also physically intuitive 
(although the price to pay for this simplicity is 
oversimplification).

Note that our toy model applies to a perturbed metric and fails if
there are large primordial black holes.

\section{Generalizing the toy model}
\label{sec:3}

We now wish to relax the unphysical assumption of spherical symmetry. The appropriate definition of quasi-local energy is given by the Hawking-Hayward construct $M_\text{HH}$ defined for closed spacelike orientable 2-surfaces by
\be
M_\text{HH}:=\frac{1}{8\pi}\sqrt{\frac{A}{16\pi}}\int_S\mu\left( \mathcal{R}+\theta_+\theta_- -\frac{1}{2} \sigma_{ab}^+ \sigma^{ab}_--2\omega_a \omega^a\right)
\ee
where $\mathcal{R}$ is the induced Ricci scalar on the 2-surface $S$, $\theta_\pm$ and $\sigma_{ab}^\pm$ are the expansions and shear tensors of a pair of null geodesic congruences (outgoing and ingoing from the surface $S$), $\omega^a$ is the projection onto $S$ of the commutator of the null normal vectors to $S$, $\mu$ is the volume 2-form on $S$ and $A$ is the area of $S$ \cite{HawkingHayward}. 

We wish to calculate $M_\text{HH}$ for the metric \eqref{CNgauge} with $\phi_N$ and $\psi_N$ general functions of all the coordinates and attempt to decompose it into a `local' and a `cosmological' piece. Working within general relativity, so that $\psi_N=\phi_N$, we will do this in two steps: firstly we will calculate $M_\text{HH}$ for the post-Newtonian metric \eqref{postNewtonianmetric} and then make use of a known transformation property of $M_\text{HH}$ under conformal transformations to extend the result to the conformally related metric \eqref{CNgauge} but without the assumption of spherical symmetry.  We will also calculate $M_\text{HH}$ directly from the metric \eqref{CNgauge} which will allow us a finer decomposition of the local part of $M_\text{HH}$ into a purely Weyl (vacuum) and a local piece coming from the perturbation itself.

\subsection{Non spherically-symmetric post-Newtonian metric}

The metric \eqref{postNewtonianmetric} is a perturbation $g_{ab}=g_{ab}^{(0)}+\delta g_{ab}$ of Minkowski spacetime $g_{ab}^{(0)}$ for which $\sigma_{ab}^\pm=0$ and $\omega^a=0$ for any 2-surface. Therefore we must have, for the post-Newtonian metric,
\begin{align}
\sigma_{ab}^\pm&=\mathcal{O}\left(\phi_N\right) \,,\\
\omega^a&=\mathcal{O}\left(\phi_N\right) \,,
\end{align}
hence the squared contributions to $M_\text{HH}$ from these tensors are $\mathcal{O}\left(\phi_N^2\right)$. That is, to first order we have
\be
M_\text{HH}:=\frac{1}{8\pi}\sqrt{\frac{A}{16\pi}}\int_S\mu\left( \mathcal{R}+\theta_+\theta_- \right) \,.
\ee

Now, from the fully general contracted Gauss equation
\be\label{Gausscontracted}
\mathcal{R}+\theta_+\theta_--\sigma_{ab}^+ \sigma^{ab}_-= h^{ac}h^{bd}R_{abcd}\, , 
\ee
where $h_{ab}$ is the induced metric on the 2-surface $S$, we compute explicitly for the post-Newtonian metric 
\begin{align}
\mathcal{R}+\theta_+\theta_-&=\left(h^{ac}_{(0)}+\delta h^{ac}\right)\hspace{-1mm} \left(h^{bd}_{(0)}+\delta h^{bd}\right)\hspace{-1mm}R_{abcd} \notag \\
&=h^{ac}_{(0)}h^{bd}_{(0)} R_{abcd} +\mathcal{O}\left(\phi_N^2\right)
\end{align}
showing this integrand to be of order $\mathcal{O}\left(\phi_N\right)$ where we have used the fact that the Riemann tensor is first order in the potentials.  The ``background'' inverse induced metric $h^{ab}_{(0)}$ is diagonal, 
\be
h^{ab}_{(0)}=\delta^a_\theta \delta^b_\theta r^{-2}  + \delta^a_\phi \delta^b_\Phi r^{-2} \sin^{-2}\theta
\ee
whence
\be
 \mathcal{R}+\theta_+\theta_-=2\, \frac{ R_{\theta\phi\theta\phi}}{r^4\sin^2\theta}.
\ee
Using an algebraic software tool such as {\sf Maple} it is simple to compute this Riemann component. After discarding all products of $\phi_N$ with any derivative $\phi_N{}_{,\mu}$ and second order terms $\phi_N^2$, $\phi_N{}_{,\mu}^2$, one is left with
\begin{eqnarray}
R_{\theta\phi\theta\phi} &=& r^2 \cos\theta\sin\theta \, \frac{\partial \phi_N}{\partial \theta}+2r^3 \sin^2\theta \frac{\partial \phi_N}{\partial r} \nonumber \\
&&\nonumber\\
&\, & +r^2\sin^2\theta \frac{\partial^2 \phi_N}{\partial \theta^2}+ r^2\frac{\partial^2 \phi_N}{\partial \phi^2} \,.
\end{eqnarray}
Therefore, the final result for $M_\text{HH}$ is given by
\begin{eqnarray}
M_\text{HH}&=&\frac{1}{8\pi}\sqrt{\frac{A}{16\pi}}\int_S\mu \left( \frac{2}{r^2}\frac{\cos\theta}{\sin\theta}\frac{\partial \phi_N}{\partial \theta} +\frac{4}{r}\frac{\partial\phi_N}{\partial r}\right.\nonumber \\
&&\nonumber\\
&\, & \left. +\frac{2}{r^2}\frac{\partial^2 \phi_N}{\partial \theta^2}+ \frac{2}{r^2\sin^2\theta} \frac{\partial^2 \phi_N}{\partial \phi^2} \right)\,. \label{E:MHH}
\end{eqnarray}

As a check, assume that $\phi_N=\phi_N(r)$ only. Then the first, third, and fourth terms in the bracket of \eqref{E:MHH} vanish and the second term can be brought out of the integral leaving
\begin{align}
M_\text{HH}&=\frac{1}{8\pi}\sqrt{\frac{A}{16\pi}}\frac{4}{r} \frac{d\phi_N}{dr} \int_S \mu \notag\\
&=\left(\frac{A}{\pi}\right)^{3/2}\frac{1}{8r}\frac{d\phi_N}{dr}\notag\\
&=R^2\frac{d\phi_N}{dr}\notag\\
&=M_\text{MSH}
\end{align}
as required (recall the result in Eq.~\eqref{E:ode}), where we have made use of $R=r+\mathcal{O}\left(\phi_N\right)$ from \eqref{arealradius} in the second to last line. 

In a subsequent section we will be interested in the question of gauge invariance for cosmological perturbations. To this end we note that the Riemann tensor is decomposed into Ricci and Weyl parts
according to \cite{Wald} 
\be
R_{abcd} = g_{a[c}R_{d]b} - g_{b[c}R_{d]a} -\frac{R}{3}\, 
g_{a[c} g_{d]b} + {C_{abcd}}   \label{decomposeRiem} \;\;
\ee
and, accordingly, the Hayward-Hawking mass is decomposed
into Ricci and Weyl parts
\begin{align} \label{Mdecomposition}
M_\text{MSH}&= M_\text{Ricci}+M_\text{Weyl}\notag \\
&=\frac{1}{8\pi}\sqrt{\frac{A}{16\pi}} \int_S\mu\,\left(\frac{4}{3}\nabla^2\phi_N+ h^{ac}h^{bd}C_{abcd} \right)
\end{align}
where $\nabla^2$ is the Laplacian. This is a clean and sharp decomposition, which is 
gauge-independent (we will show this below in Sec.~\ref{S:gauge}).

\subsection{Non spherically-symmetric post-FLRW metric}

\subsubsection{Calculation of $M_\text{HH}$ via conformal transformation}

Using the result \eqref{E:MHH} and a previous result for the transformation of the Hawking-Hayward energy under conformal transformations \cite{PVFLL}, we can compute the Hawking-Hayward energy for the post-FLRW metric \eqref{CNgauge}. The general result for the conformal transformation of $M_\text{HH}$ is 
\begin{eqnarray}
\tilde{M}_{\text{HH}} &=& \sqrt{\frac{\tilde{A}}{A}}M_{\text{HH}}+\frac{1}{4\pi} \sqrt{ \frac{\tilde{A}}{16\pi}}
\int_{S} \mu \left[ h^{ab} 
\left( \frac{
2 \nabla_a \Omega \nabla_b  \Omega}{\Omega^2}
 \right.\right.\nonumber \\
& \, & \left.\left. - \frac{\nabla_a \nabla_b \Omega}{\Omega} \right)- \frac{ \nabla^c  \Omega \nabla_c
\Omega}{\Omega^2} \right] \label{conh} \, , 
\end{eqnarray}
where the metric $h_{ab}$ and measure $\mu$, as well as the covariant derivatives $\nabla_a$ are those with respect to the conformally related post-Newtonian metric \eqref{postNewtonianmetric}. In the special case we are interested in, $\Omega=\Omega(\eta)$ only and hence the first term inside the integrand vanishes. The second term is computed using
\begin{align}
h^{ab}\nabla_a\nabla_b\Omega&= h^{ab}\nabla_a \left( \delta_b^\eta\,\Omega_{,\eta} \right) \notag\\
&=h^{ab}\left(\delta_a^\eta\,\delta_b^\eta\,\Omega_{,\eta}^2+ \Gamma^{\eta}{}_{ba}\,\Omega_{,\eta}\right) \notag \\
&=\left(\Gamma^\eta{}_{\theta\theta}h^{\theta\theta} +\Gamma^{\eta}{}_{\phi\phi}h^{\phi\phi}\right)\Omega_{,\eta}\notag \\
&=-\frac{2\phi_N{}_{,\eta}}{\left(1+2\phi_N\right)2}\Omega_{,\eta}\notag \\
&\simeq-2\phi_N{}_{,\eta}\Omega_{,\eta} 
\end{align}
where we truncated to first order in the perturbation in the last line. The last term in the integrand is simpler where we have
\begin{align}
\nabla_a\Omega \nabla^a \Omega&=g^{\eta\eta}\Omega_{,\eta}^2 \notag \\
&=-\frac{\Omega_{,\eta}^2}{\left(1+2\phi_N\right)}\notag \\
&=-\Omega_{,\eta}^2+2\Omega_{,\eta}^2\phi_N \,.
\end{align}
Therefore, the final result is 
\begin{eqnarray}
\tilde{M}_\text{HH}&=&
\Omega M_\text{HH}+\frac{R\Omega_{,\eta}}{4\pi}\left( \int_S\mu \, \phi_N{}_{,\eta} -\frac{\Omega_{,\eta}}{\Omega}\int_S\mu \, \phi_N \right) \nonumber\\
&&\nonumber\\
&\, & +\frac{R^3}{2} \frac{\Omega_{,\eta}^2}{\Omega}\, . \label{E:HH_tilde}
\end{eqnarray}
At late times in the dust-dominated era, $\Omega(\eta)=\Omega_0\eta^2$ and the right hand side is saturated by the first term, which is constant since the second term decays as $\eta^{-1}$. That is, at late cosmological times
\be
\tilde{M}_\text{HH}\simeq \Omega M_\text{HH} \,.
\ee

Again, we can perform a check of the general result \eqref{E:HH_tilde} under the assumption that $\phi_N=\phi_N(r)$ only. Writing $a:=\Omega$ for comparison,  our formula \eqref{E:HH_tilde} simplifies in that case to
\begin{align}
\frac{\tilde{M}_\text{HH}}{a}&=M_\text{HH} +\frac{R}{4\pi} \mathcal{H}\left(\mathcal{H}\phi_N A\right)+\frac{R^3}{2} \mathcal{H}^2 \notag \\
&= M_\text{HH}+\frac{R^3 \mathcal{H}^2}{2}\left(1-2\phi_N\right)\, ,
\end{align}
which agrees exactly with the formula \eqref{E:decompsition} written in terms of the areal radius in the post-Newtonian space $R=\tilde{R}/a$.

\subsubsection{Direct calculation of $M_\text{HH}$ from the metric \eqref{CNgauge}}

An independent calculation of the result \eqref{E:HH_tilde} can be performed directly by making use of the formula \eqref{Gausscontracted}.

The metric \eqref{CNgauge} is a perturbation of the spatially flat FLRW spacetime for which $\omega_a$ vanishes \cite{HawkingHayward}. This implies that $\omega_a=\mathcal{O}\left(\phi_N\right)$ for the metric \eqref{CNgauge} and hence the squared contribution $\omega_a\omega^a$ can be neglected in the computation of $\tilde{M}_\text{HH}$. Therefore the left hand side of \eqref{Gausscontracted} coincides with $\tilde{M}_\text{HH}$ up to first order in $\phi_N$. The right hand side of that equality is easy to compute (using {\sf Maple} for example) for the metric $\tilde{g}$ given in \eqref{CNgauge}.   

We wish to decompose the $\tilde{M}_\text{HH}$ into `local' and `cosmological' contributions, similarly to what we did in the spherically symmetric case. Using again the general decomposition \eqref{decomposeRiem} it can be shown that to first order
\begin{align}
\tilde{h}^{ac}\tilde{h}^{bd}\tilde{R}_{abcd}&=\tilde{h}^{ac}\tilde{h}^{bd}\tilde{C}_{abcd}+\frac{4}{3\Omega^2} \nabla^2\phi_N \notag \\
&-\frac{4}{\Omega^2}\left(\frac{\Omega_{,\eta}}{\Omega}\right)^2 \phi_N-\frac{4}{\Omega^2}\left(\frac{\Omega_{,\eta}}{\Omega}\right)\phi_N{}_{,\eta} \notag \\
&+\frac{2}{\Omega^2}\left(\frac{\Omega_{,\eta}}{\Omega}\right)^2
\end{align}
where $\nabla^2$ is the spatial Laplacian in spherical coordinates for the post-Newtonian metric $g_{ab}$.  Therefore the full Hayward Hawking energy is given by
\begin{align}
\tilde{M}_\text{HH}&= \frac{\Omega R}{16\pi}\int_S\tilde{\mu}\left( \tilde{h}^{ac}\tilde{h}^{bd}\tilde{C}_{abcd}\right)+\frac{\Omega  R}{12\pi}\int_S \mu \left(\nabla^2\phi_N\right) \notag \\
&- \frac{R}{4\pi}\Omega_{,\eta} \left( \int_S\mu \, \phi_N{}_{,\eta} +\frac{\Omega_{,\eta}}{\Omega}\int_S\mu \,\phi_N \right)+\frac{R^3}{2} \frac{\Omega_{,\eta}^2}{\Omega}\, . \label{E:direct}
\end{align}

Each of the terms in \eqref{E:direct} have a clear physical interpretation. The first two terms are `local', the first of gravitational origin being purely Weyl while the second being the contribution from the perturbation itself. Indeed, using the Poisson equation this second term can be written as
\be
\frac{\Omega R}{12\pi}\int_S \mu \,\nabla^2\phi_N = \frac{\Omega R}{3}\int_S \mu \,\rho_\phi
\ee
showing it to be simply the integrated density $\rho_\phi$ of the perturbation itself. The third, fourth and fifth terms in \eqref{E:direct} are `cosmological', the fifth being the mass energy of the cosmological fluid (responsible for the cosmological expansion) contained in the sphere of radius $R$. By the Friedman equation this fifth term can be written
\be
\frac{R^3}{2} \frac{\Omega_{,\eta}^2}{\Omega}=\Omega\,\text{Vol}\,\frac{3}{8\pi}H^2=\Omega \text{Vol} \, \rho_\text{cos}
\ee
where $H=\Omega_{,\eta}/\Omega^2$ is the Hubble parameter and $\rho_\text{cos}$ is the density of the (unperturbed) cosmic fluid. 

Due to the conformal invariance of the $\left( \begin{array}{c} 1 \\
3  \end{array} \right)$ Weyl tensor ${C_{abc}}^d$, we can relate the pure Weyl term to the corresponding term computed with respect to the post-Newtonian metric by
\be
\tilde{h}^{ac}\tilde{h}^{bd}\tilde{C}_{abcd}=\frac{1}{\Omega^2}\,  h^{ac}h^{bd}C_{abcd}
\ee
which leads us to re-derive the result \eqref{E:HH_tilde}
\begin{eqnarray}
\tilde{M}_\text{HH}& = & \Omega M_\text{HH} - \frac{R}{4\pi}\Omega_{,\eta} \left( \int_S\mu \, \phi_N{}_{,\eta} +\frac{\Omega_{,\eta}}{\Omega}\int_S\mu \,\phi_N \right)\notag \\
&&\nonumber\\
& \, & +\frac{R^3}{2} \frac{\Omega_{,\eta}^2}{\Omega}\,. 
\end{eqnarray}
where we recall the result \eqref{Mdecomposition}.

This alternative calculation through the use of the Weyll tensor allows us the additional refinement of the local contribution, which is the only contribution to $\tilde{M}_\text{HH}$ which survives at late times during dust domination, into purely Weyl (vacuum) and purely Ricci (matter) parts as
\begin{align}
\tilde{M}_\text{HH}& \rightarrow \frac{\Omega R}{16\pi}\int_S\tilde{\mu}\left( \tilde{h}^{ac}\tilde{h}^{bd}\tilde{C}_{abcd}\right)+\frac{\Omega  R}{12\pi}\int_S \mu \left(\nabla^2\phi_N\right) \notag \\
&=\Omega M_\text{HH} \label{E:limit}
\end{align}
These are the first two terms in the full result \eqref{E:direct}.  

This result corroborates the spherically symmetric case where we saw that the cosmological part of the quasi-local energy became negligible at late times in the dustdominated era.

\subsection{Gauge invariance} \label{S:gauge}

One of the advantages of this work over other approaches in the literature (for example the works of Refs.~\cite{ChisariZaldarriaga, GreenWald, Adamek})  on the question of the Newtonian approximation in large scale structure formation is that our results are 
explicitly gauge-invariant, as well as being physically transparent. Below we illustrate 
this gauge invariance for both the general and spherically symmetric decompositions 
of the quasi-local energy to first order in perturbation theory.

Firstly, it is known that $M_\text{HH}$, and hence the special case $M_\text{MSH}$, is a 
gauge-invariant quantity. As we have already noted, to first order in perturbations 
the quasi-local energy $M_\text{HH}$ is proportional to the surface integral of the 
scalar quantity
\be
\mathcal{R}+\theta_+\theta_--\sigma_{ab}^+ \sigma^{ab}_-= h^{ac}h^{bd}R_{abcd}.
\ee
This integrand splits naturally into two parts based on the gauge-invariant splitting 
of the Riemann tensor into a trace (`Ricci') and tracefree (Weyl) component as we have already noted:
\be
R_{abcd} = g_{a[c}R_{d]b} - g_{b[c}R_{d]a} -\frac{R}{3}\, 
g_{a[c} g_{d]b} + {C_{abcd}} \;.
\ee
At late times $M_\text{HH}$ converges to precisely the Weyl contribution and an extra 
piece, which sums the energy density of the perturbation itself (in a `quasi-local' fashion, 
see Eq.~\eqref{E:limit}). The Weyl contribution is considered `local' since it does not 
contain information about the cosmic expansion (apart from a pre-factor) while the 
additional energy arising from the perturbation, while being part of the `Ricci' 
contribution to $M_\text{HH}$, is clearly local in character. The (first order) 
gauge invariance arises from the fact that Eq.~\eqref{E:limit} relates a gauge-invariant
quantity, $\tilde{M}_\text{HH}$, with two objects, one of which (the purely Weyl contribution) 
is gauge-invariant.  Therefore, to this level of perturbation theory, also the second 
term is gauge-invariant and hence the decomposition into local and cosmological parts is 
gauge-invariant at late times. Our conclusion that the local part saturates the 
quasi-local energy at late times in a dust-dominated era is also a gauge-invariant statement.

\section{Conclusions}
\label{sec:4}

Since all our understanding of the formation of large scale structures is based on Newtonian simulations, it is important to check whether using what are basically the wrong equations leads to errors and a misleading picture. Confidence based on physical intuition while running the simulations is not a substitute for careful checking. It is only recently that the works \cite{ChisariZaldarriaga} and \cite{GreenWald} have taken on this problem and reassured us about the validity of the final results of these Newtonian calculations even if the wrong premises are assumed. In view of the importance of this subject, however, it seems that more approaches with different techniques should be pursued. The new approach that we present here consists of splitting the Hawking-Hayward quasi-local energy of a closed, orientable, spacelike 2-surface into a ``Newtonian'' part due entirely to the perturbations of a FLRW space (which later become the large scale structures), and a ``cosmological'' part which contains the relativistic effects. The idea is that, if Newtonian physics is really coming to dominate and the relativistic effects remain small, then the physical mass-energy of a region of spacetime should be dominated by a Newtonian contribution while the relativistic effects should give a negligible contribution.
We have found and presented the relevant decomposition of the quasi-local energy; this decomposition is gauge-invariant. We have first analyzed a simple toy model to illustrate the idea and provide a sense of how the calculation proceeds, and then we have moved on to realistic situations. 
For general perturbations, it is not trivial to prove that the sought for decomposition of the quasi-local energy into Newtonian and relativistic parts exists; it follows from the geometric and gauge-invariant splitting of the Riemann tensor of the perturbed FLRW space into a Ricci part and a Weyl part. At late times in the dust era, the Newtonian part of the quasi-local mass dominates over the contributions containing the relativistic effects. The advantage of our approach (limited to linear order in the metric perturbations) is that it is covariant and gauge-invariant, yet relatively simple to follow. Our result provides confidence in the Newtonian simulations and agrees with the spirit of the analyses of \cite{ChisariZaldarriaga, GreenWald}, although the methods are quite different.

\begin{acknowledgments} 

V.F. thanks Markus Haider for interesting discussions.  
This research is supported by Bishop's University and by the 
Natural Sciences and Engineering Research Council of Canada ({\em 
NSERC}). 
\end{acknowledgments}



\end{document}